%% file: main.tex
\documentclass[%
superscriptaddress,
 amsmath,amssymb,
 aps,
 pre,
 twocolumn,
]{revtex4-1}

\usepackage{graphicx}% Include figure files
\usepackage{dcolumn}% Align table columns on decimal point
\usepackage{bm}% bold math
\usepackage{xcolor}
\usepackage{url}
\usepackage{hyperref}
\usepackage[caption=false]{subfig}
\usepackage{booktabs}
\usepackage[section]{placeins}
\makeatletter
\AtBeginDocument{%
  \expandafter\renewcommand\expandafter\subsection\expandafter{%
    \expandafter\@fb@secFB\subsection
  }%
}
\makeatother

\begin{document}

% \preprint{APS/123-QED}

\title{Robustness modularity in complex networks}

\author{Filipi N. Silva}
\affiliation{Indiana University Network Science Institute}

\author{Aiiad Albeshri}
\affiliation{Department of Computer Science, Faculty of Computing and Information Technology
King Abdulaziz University, Jeddah 21589, Kingdom of Saudi Arabia}

\author{Vijey Thayananthan}
\affiliation{Department of Computer Science, Faculty of Computing and Information Technology
King Abdulaziz University, Jeddah 21589, Kingdom of Saudi Arabia}

\author{Wadee Alhalabi}
\affiliation{Department of Computer Science, Faculty of Computing and Information Technology
King Abdulaziz University, Jeddah 21589, Kingdom of Saudi Arabia}

\author{Santo Fortunato}
\affiliation{Indiana University Network Science Institute (IUNI)}
\affiliation{Luddy School of Informatics, Computing and Engineering, Indiana University}

% \author{Filipi Nascimento Silva}
% \affiliation{Indiana University Network Science Institute, Bloomington, IN, USA.
% }


\date{\today}

\begin{abstract}
A basic question in network community detection is how modular a given network is. This is usually addressed by evaluating the quality of partitions detected in the network. The Girvan-Newman (GN) modularity function is the standard way to make this assessment, but it has a number of drawbacks. Most importantly, it is not clearly interpretable, given that the measure can take relatively large values on partitions of random networks without communities. Here we propose a new measure based on the concept of robustness: modularity is the probability to find trivial partitions when the structure of the network is randomly perturbed. This concept can be implemented for any clustering algorithm capable of telling when a group structure is absent. Tests on artificial and real graphs reveal that robustness modularity can be used to assess and compare the strength of the community structure of different networks. We also introduce two other quality functions: modularity difference, a suitably normalized version of the GN modularity; information modularity, a measure of distance based on information compression. Both measures are strongly correlated with robustness modularity, and are promising options as well.

\end{abstract}

\maketitle

%\tableofcontents

\section{\label{sec1}Introduction}

The simple network, or graph, representation of many systems in nature, society, and technology has provided invaluable insight into the structure and function of these systems~\cite{newman10,barabasi16,menczer20}. A general property of networks is their community structure, i.e. their natural division into cohesive groups of nodes, called communities, modules, or clusters, that are loosely connected to each other~\cite{girvan02}. Communities represent the building blocks of networks, and are key to understand both their structural properties and the dynamics of processes running on them. This is why community detection has become one of the most popular topics within network science~\cite{porter09,fortunato10,fortunato16}.

Detecting communities in networks is an unsupervised classification problem, and as such it is ill-defined. 
One of the main questions is how to define how ``modular" a network is. This calls for the definition of a measure that estimates the goodness of the community structure. The Girvan-Newman (GN) modularity function~\cite{newman04b} does precisely that, in principle: the quality of the community structure of a network is the largest value of modularity over all partitions of the graph.
In practice, however, GN modularity has some serious limitations. 
For example, the largest modularity value does not necessarily correspond to the most pronounced community structure, because the measure has a preferential community scale~\cite{fortunato07}. Also, the maximum modularity of partitions of random graphs without group structure, e.g. Erd\H{o}s-R\'enyi graphs~\cite{erdos59}, can still attain surprisingly high values~\cite{guimera04}. According to GN modularity, then, such random networks do have community structure, against intuition. For this reason, it may be misleading to qualify the group structure based on its GN modularity, or to establish which of two networks is more modular than the other by comparing their maximum modularity scores.

In this paper, we propose an entirely different approach to the problem of evaluating the modularity of a graph. Our null hypothesis, widely shared in the scientific community, is that, if clusters do not play a role in the generating mechanism of a network, the latter is not modular. In the example above, Erd\H{o}s-R\'enyi graphs of any size should have modularity zero. Assessing how modular a graph is then amounts to quantifying the difference between the graph and an Erd\H{o}s-R\'enyi graph with an equal number of nodes and links, from a community perspective. 
A natural way to do that is by gradually modifying the structure via random perturbations, consisting of random rewirings of a growing fraction of links, until communities are disrupted and the network becomes equivalent to an Erd\H{o}s-R\'enyi graph. Modularity is defined as the probability that, as the network is perturbed, a given clustering algorithm finds partitions different from the trivial ones detected in random graphs, like the division into a single cluster. Our definition is rooted in the notion of network robustness~\cite{karrer08}: the more modular the graph, the larger the perturbation needed to lose
the communities. This is why we call it \textit{robustness modularity} (RM). We stress that the score does not require any analysis or evaluation of the actual partitions found by the chosen clustering method; what matters is only whether they are trivial partitions or not.

The measure is properly normalized and can thus be used to rank distinct networks according to their scores, regardless of their size or kind. Furthermore, it can be computed by using any community detection technique, as long as it is able to recognize the absence of communities, though this is a test most methods fail at. As a proof of concept, in this paper, we will use a posteriori stochastic blockmodeling~\cite{peixoto20} as clustering technique, due to its remarkable ability to identify random graphs. The analysis of artificial benchmark graphs and real networks shows that RM is a reliable criterion to assess and compare the quality of partitions. Unsurprisingly, RM is poorly correlated with GN modularity, which reaffirms the concerns on the use of the latter. However, RM is strongly correlated to an intuitive normalization of GN modularity, as well as to a function describing the distance between the network and its Erd\H{o}s-R\'enyi ensemble in terms of information content.

\section{Methods}

\begin{figure}[!h]
\centering
    \includegraphics[width=\columnwidth]{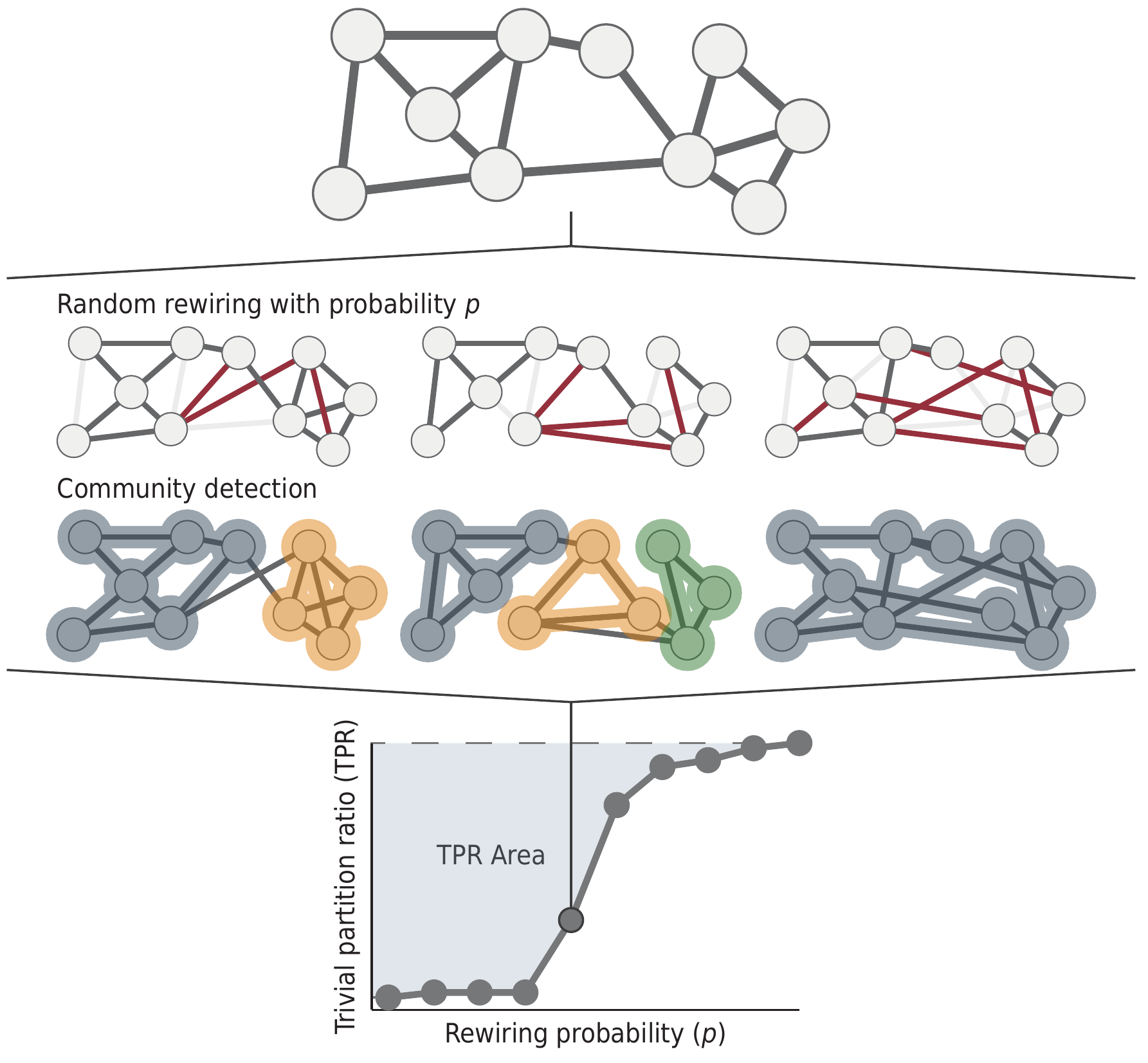}
    \caption{Schematic representation of the proposed methodology. The network is the one at the top. First, links in the network are randomly rewired with a probability $p$. Then, a clustering algorithm is applied on all perturbed networks, producing a TPR profile (bottom). The RM is the area above the curve of the TPR profile (highlighted). }
\label{fig:TPR_scheme}
\end{figure}

Figure~\ref{fig:TPR_scheme} shows a schematic description of our procedure. There are three main ingredients:
\begin{enumerate}
    \item The graph perturbation method;
    \item The clustering algorithm used to detect the communities of the perturbed network;
    \item The calculation of the RM.
\end{enumerate}

There are different ways to perturb the structure of a graph. We opted for a simple unconstrained rewiring, in that each link is rewired with a probability $p$, and the endpoints of the rewired link can be any two nodes of the network. Therefore, $p$ is the expected fraction of rewired links. For $p=1$ we just recover an Erd\H{o}s-R\'enyi random graph with the same number of nodes and links of the original network. The network is perturbed $25$ times for each fraction $p$.

We have also tried an alternative popular choice, in that links are rewired such that the degree sequence of the network is preserved. Such procedure has the problem that structures, where hubs are connected to each other may be very stable as the network is randomized, as there might be no other options for the hubs to keep their degree. This would confer such networks an artificially high modularity.

In principle, we could use any method to detect partitions on the perturbed network. However, popular choices like the Louvain algorithm~\cite{blondel08} and Infomap~\cite{rosvall08} are unable to recognize random graphs, which is a major caveat on their applicability. Therefore, after some preliminary tests with these techniques, we decided to go for statistical inference algorithms based on \textit{stochastic blockmodels} (SBM)~\cite{fienberg81, holland83, wasserman87, peixoto20}, which are especially good at detecting random graph configurations. In particular, we infer a microcanonical SBM: the best partition is the one minimizing the total amount of information required to describe the network and the model, expressed by the \textit{description length}~\cite{peixoto13}. We adopt a version of SBM with degree correction~\cite{karrer11}, that generates graphs with the same degree sequence of the network at study. The description length for this model is defined in Appendix B. We used the Graph-tool v2.35~\cite{peixoto_graph-tool_2014} implementation of degree-corrected SBM. This algorithm delivers the trivial partition $T$ into a single cluster when it analyzes a random graph without group structure. For each perturbed network, we run the algorithm just once, therefore we obtain $25$ partitions in total for each value of $p$.

For each value of the rewiring probability $p$ we define the \textit{trivial partition ratio} (TPR) as the fraction of trivial partitions $T$ found among all perturbed graph configurations for that value of $p$:

%\begin{widetext}
\begin{equation}
    TPR(p)=\frac{\mbox{\# perturbed graph configurations with \textit{T}}}{\mbox{\# perturbed graph configurations}}.
\end{equation}
%\end{widetext}

When $TPR(p)$ is equal to zero, the algorithm does not find trivial partitions. When $TPR(p)$ is equal to one, instead, all divisions are trivial. Therefore, $TPR(p)$ is the probability that there is no community structure for a given $p$.
In Fig.\ref{fig:TPR_scheme} (bottom) we show a typical profile of the TPR as a function of $p$.
 If the graph has community structure, for small perturbations ($p\sim 0$) we expect that the TPR is close to zero, i.e. that we never find trivial divisions in the perturbed graphs, as small perturbations should not appreciably modify the communities. On the other hand, for sufficiently strong perturbations, the community structure is destroyed and the network becomes equivalent to a random graph with no groups, so the TPR gets close to one. 
 
 Therefore, the area above the TPR curve (highlighted in the figure) is a measure of the strength of the modular structure of the graph. This area quantifies our RM, and we indicate it as $Q_r$. Since both $p$ and the TPR range between zero and one, so does $Q_r$, with zero indicating the absence of community structure and one the strongest possible clustering.

\section{Results}

\subsection{Model Networks}
\label{sec:modnet}

Artificial networks with built-in communities are regularly used to validate clustering algorithms~\cite{fortunato10}.
Here we use the standard SBM, which generates networks divided into $q$ groups of identical size $N_c$, and nodes with approximately the same degree. For our numerical experiments we used $q=4$, $N_c=1250$ and the average degree of the nodes is $\langle k\rangle=10$. 
The mixing parameter $\mu$ is the average ratio between the number of neighbors that a node has in communities other that its own, and the degree of the node. Hence, $\mu$ indicates how pronounced the community structure is. If $\mu=0$ clusters are disjoint from each other, i.e., all links fall within communities, and are easily detectable.  In this scenario we deal with purely \textit{assortative} communities. For $\mu=1$ links fall exclusively between groups and communities are purely \textit{disassortative}. Normally $\mu$ is varied to explore different strengths of communities.

\begin{figure}[!h]
    \centering
    \includegraphics[width=\columnwidth]{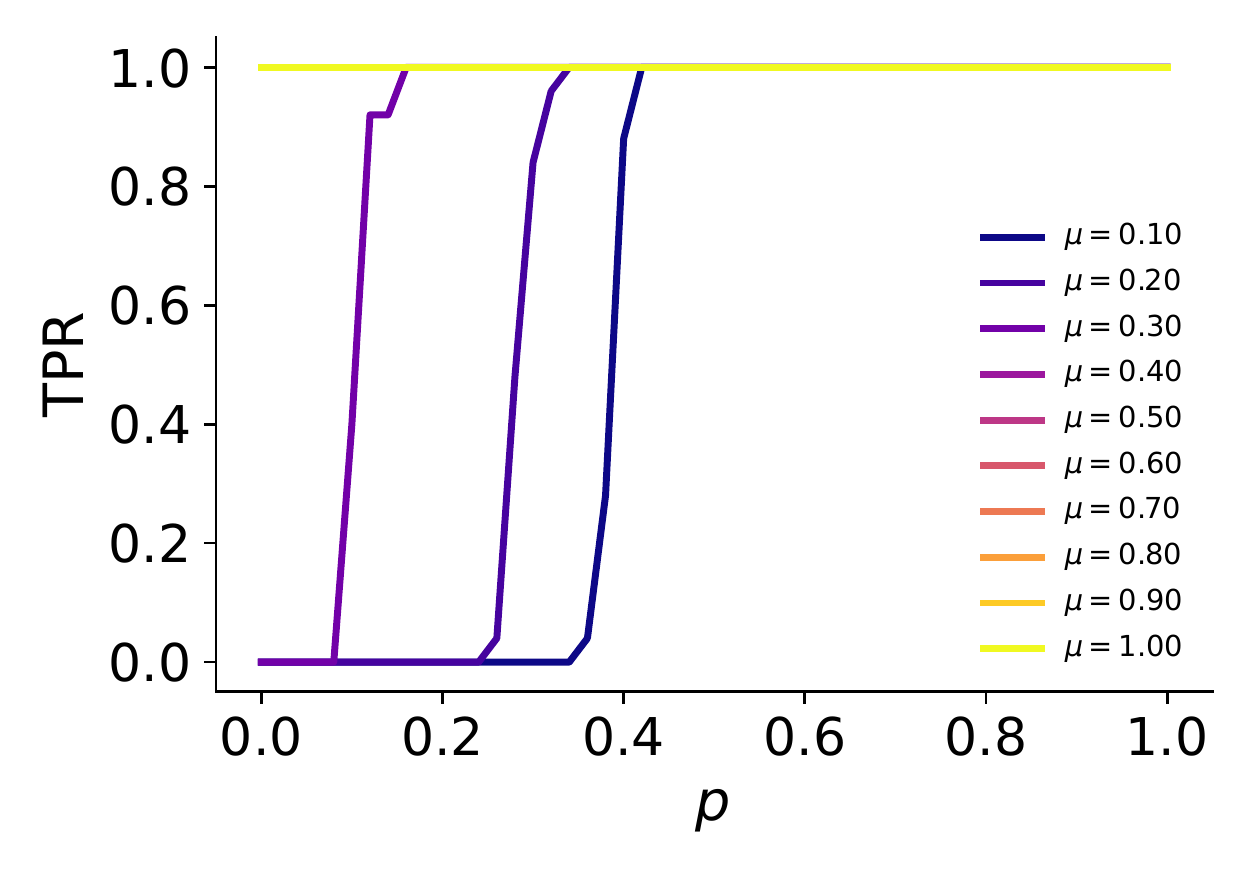}
    \caption{Trivial partition ratio (TPR) for SBM benchmark graphs. }
    \label{fig:TPR_Benchmarks}
\end{figure}

In Fig.~\ref{fig:TPR_Benchmarks} we report the TPR profiles for graphs generated by the standard SBM and distinct $\mu$-values. 
We see that there is a threshold behavior, in that the TPR is zero until a certain amount of noise $p$, after which it jumps to one. This is interesting, as it indicates that the transition from a network with communities to its perturbed group-free counterpart happens quickly, for a certain amount of noise, as in phase transitions. This is due to the fact that communities have approximately the same edge density by construction, hence they are destroyed at about the same time. In the range of $p$ where the TPR plateaus to one, the graph is virtually indistinguishable from an Erd\H{o}s-R\'enyi random graph with the same number of nodes and links. As expected, the higher the $\mu$, the earlier the transition occurs, signaling a lower modularity $Q_r$.

For $\mu\gtrsim 0.4$, ${TPR}(p)=1$ for all values of $p$, indicating that the model graph has no communities ($Q_r=0$). 
So, from the point of view of community structure, graphs generated via the standard SBM with $\mu\gtrsim 0.4$ ($q=4$, $\langle k \rangle =10$) are equivalent to classic Erd\H{o}s-R\'enyi random graphs. This is actually unexpected, because, as $\mu$ increases, after a range in which there are no detectable groups, we should switch to disassortative community structure, which is supposed to be detectable by a technique based on SBMs like the one we have used. Instead, even for $\mu=1$, the TPR profile is totally flat, starting from the unperturbed graph ($p=0$). In Appendix C we will show that the planted disassortative partition is indeed worse than the trivial division into one group. While it would be ideal to assess the robustness of different types of community structure, we can settle for the robustness of assortative communities, which are the focus of most graph clustering analyses. 

\begin{figure}[!h]
    \centering
    \includegraphics[width=\columnwidth]{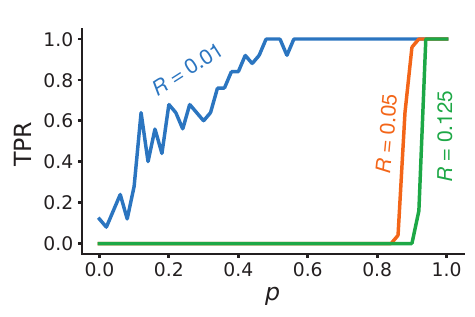}
    \caption{TPR curves obtained for the hybrid model consisting of an ER graph and a clique including a fraction R of the nodes.}
    \label{fig:TPR_Hyb_Model}
\end{figure}

We have also performed tests on the more realistic LFR benchmark graphs~\cite{lancichinetti08}, as well as on other random graph models (Appendix~\ref{apx_models}). 

\begin{figure*}[htb]
    \centering
    \includegraphics[width=\textwidth]{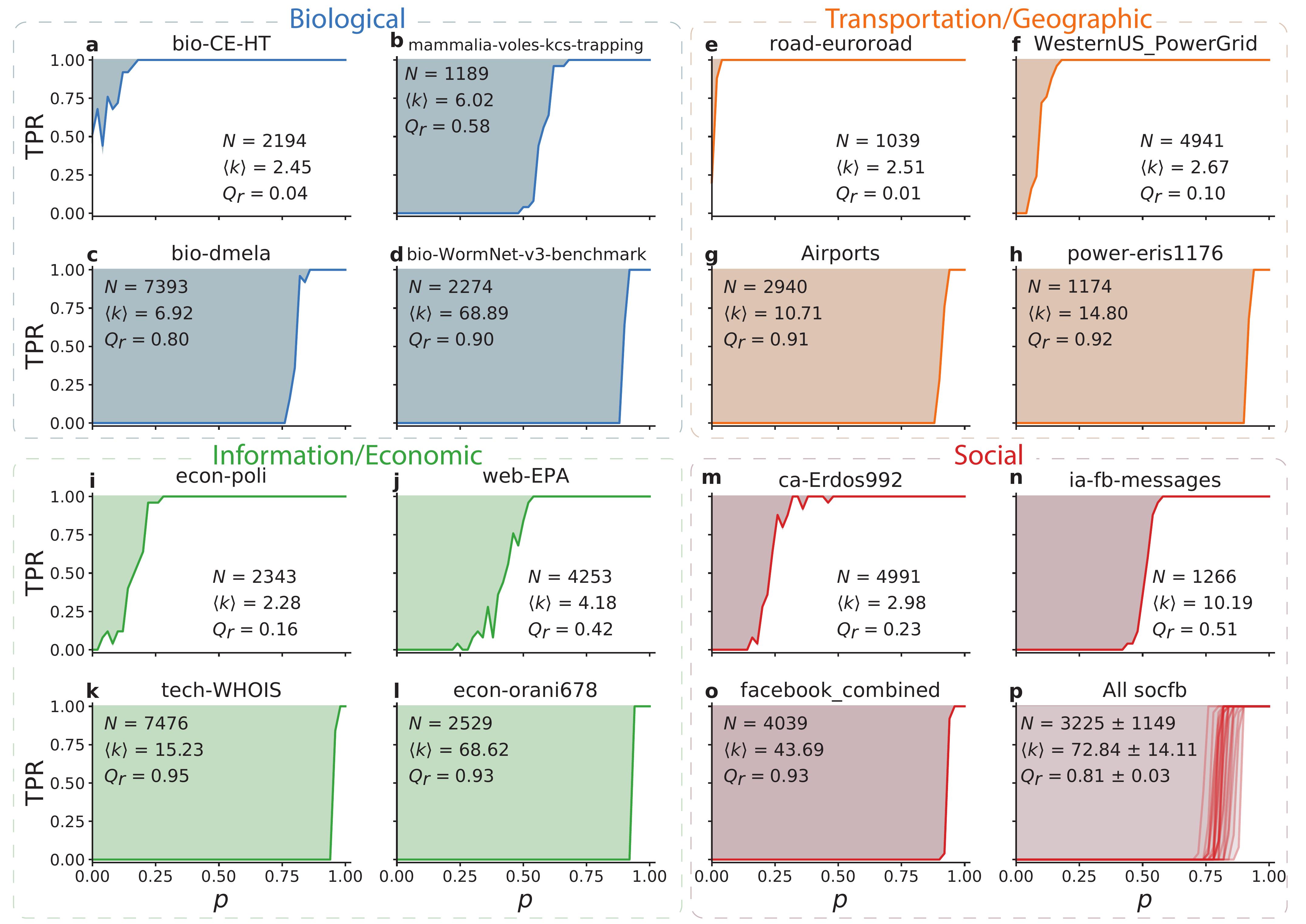}
    \caption{Typical TPR curves obtained for real networks. Number of nodes ($N$), average node degree $\langle k\rangle$ and the RM $Q_r$ are shown for each pattern. The last panel (p) shows all the TPR curves obtained from the Facebook social networks in the dataset, which are fairly similar to each other.} %will change Girvan-Newman to SBM
    \label{fig:TPR_Real}
\end{figure*}

In general, the communities of a network partition may have a highly heterogeneous structure, so they might require distinct amounts of noise (values of $p$) to be disrupted. On the other hand, as long as there is at least one cluster left, the TPR will have a value lower than one and the perturbed 
network will be deemed as modular, to some extent. Therefore, $Q_r$ actually evaluates the robustness of the strongest cluster of the partition, the last one to be disrupted by rewiring. This could potentially be a problem.
The strongest possible cluster is a clique, a subgraph whose nodes are all connected to each other. If a network had a small clique, but is otherwise weakly modular, we would be inclined to assign a low modularity score to it. However, $Q_r$ might still be large, because the clique is hard to destroy. To check for that, we consider a special model network, consisting of a clique including a fraction $R$ of the nodes, whereas the rest form an Erd\H{o}s-R\'enyi random graph. The network is built upon a random graph, by randomly choosing $NR$ nodes and adding all the necessary links among them such to form a clique.

% The total number of nodes is XXX. Number of nodes will be the same %WE MIGHT CONSIDER ACCOUNTING FOR THE FRACTION OF linkS OF THE CLIQUE, INSTEAD OF nodes. F: In that case the Ratio correspond approximately to (NR*(NR-1)/2) /(k*(N)/2+(NR*(NR)/2))
% $R=0.010$: $0.89\%$
%[???] calculate it for R=0.020 
% $R=0.050$: $19.7\%$
% $R=0.125$: $60\%$
In Fig.~\ref{fig:TPR_Hyb_Model} we show TPR profiles for three different clique sizes, including $1\%$ ($0.9\%$ of links), $5\%$ ($19.7\%$ of links), and $12.5\%$ ($60\%$ of links) of the nodes of the graph, respectively. For $R=1\%$ $Q_r$ is relatively low, while it becomes important for $R\geq 5\%$, i.e. when the clique includes an important fraction of the links. Since it is very uncommon to have such large cliques in real networks, we can safely claim that the $Q_r$ is not going to be determined by the structure of a tiny portion of the system.  

\subsection{Real Networks}

We analyzed many real datasets, from the Index of Complex Networks collection~\cite{clauset_tucker_sainz_2016}. They cover different domains: social, biological, information, economic and transportation networks. We limited the scope of the analysis to networks having between $1,000$ and $10,000$ nodes. We also removed bipartite networks and those with negative weights. For all the networks we performed the analysis only on their giant connected components. In Fig.~\ref{fig:TPR_Real} we show the TPR profiles for a selection of datasets. The full results are reported in Appendix~\ref{apx_TPR_ALL}.
We recover the same threshold behavior that we have observed for the artificial benchmarks. Inhomogeneities of the structure are reflected by small irregularities of the curves, though the latter can usually be reduced by increasing the number of perturbations for a given $p$. The transition from zero to one of the TPR sometimes is not so abrupt, which is probably due to the fact that clusters may have different robustness, so they are disrupted progressively, over a range of values for $p$.
\begin{figure*}[htb]
    \centering
    \includegraphics[width=0.95\textwidth]{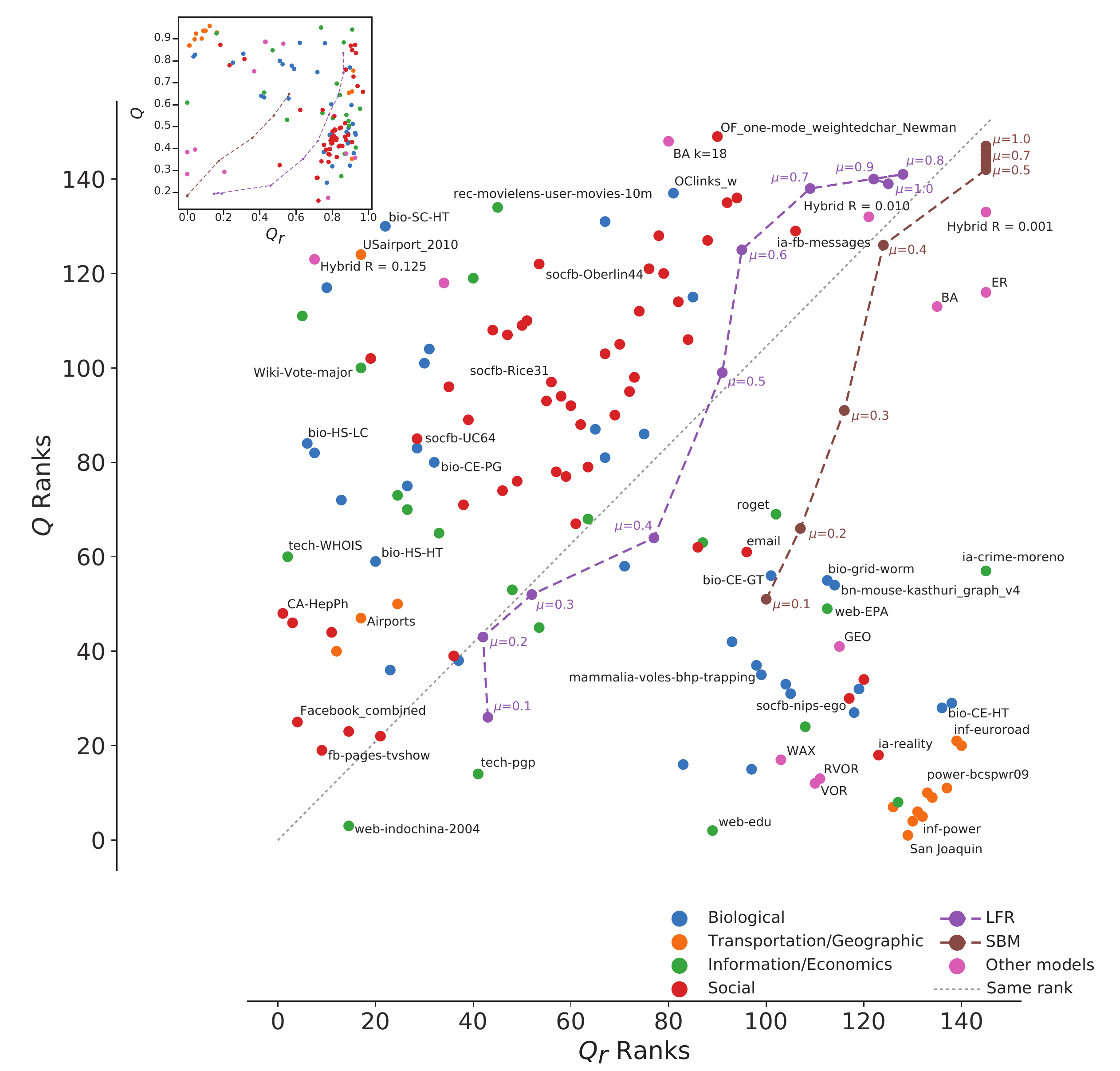}
    \caption{Relationship of the ranks of our artificial and real network datasets, obtained by using the RM $Q_r$ and the GN modularity $Q$. The Spearman correlation coefficient is $-0.12$. The Pearson correlation coefficient between the scores is $-0.03$. Inset: scatter plot between the actual values of $Q_r$ and $Q$.} 
    \label{fig:TPR_Modularity}
\end{figure*}
In Tables~\ref{table:biological}, \ref{table:transportation_geographic}, \ref{table:information_economy}, and \ref{table:social} in Appendix D we report the values of $Q_r$ for all networks we analyzed, divided by categories. We see that there is no systematic trend across categories. For instance, we cannot generally claim that social networks are more modular than biological networks, or vice versa, as the values of $Q_r$ span most of its range for both classes of graphs. It is remarkable that a few networks, like \emph{bio-CE-HT}, representing functional associations between genes in {\it C. elegans}~\cite{cho2014wormnet}, have $Q_r$ very close to zero, indicating the absence of modules. However, if we look at the structure of the network, illustrated in Figs.~\ref{fig:TPR_ModularityNorm}, \ref{fig:TPR_MDL}, and \ref{fig:MDL_ModularityNorm}, we notice that it is essentially tree-like, hence it does not appear to be modular.

\subsection{Related measures}

We may wonder how $Q_r$ relates to GN modularity $Q$. In Fig.~\ref{fig:TPR_Modularity} we compare the two measures. The maximum $Q$-value of a network is estimated by the largest value of $Q$ obtained from $100$ partitions found via the Louvain algorithm \cite{blondel08}. 
\begin{figure*}[htb]
    \centering
    \includegraphics[width=0.95\textwidth]{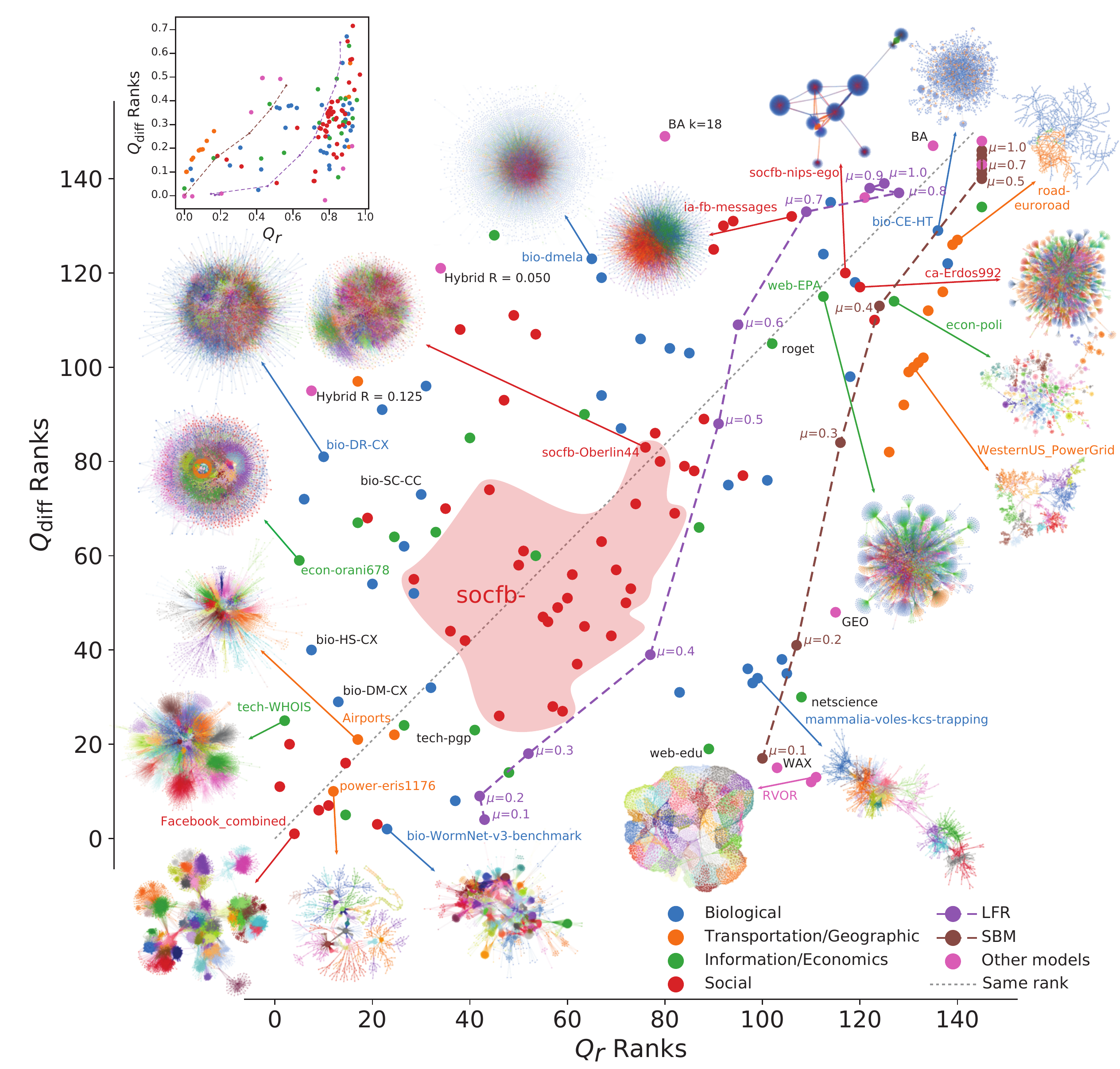}
    \caption{Relationship between $Q_{r}$ and $Q_{diff}$.
    As in Fig.~\ref{fig:TPR_Modularity} points are identified by the ranks of the corresponding datasets according to the two scores.
    The networks shown in Fig.~\ref{fig:TPR_Real} and a few others are highlighted and their respective structures are displayed. The colors of the nodes in each network identify the communities obtained from the microcanonical SBM inference. The shaded red region highlights  the social networks from the set \textit{socfb}. We connect the dots representing LFR and (standard) SBM model graphs to show that both modularity scores decrease (ranks increase) when the mixing parameter  $\mu$ increases. Inset:  scatter plot between the actual values of $Q_r$ and $Q_{diff}$.
    Spearman ($S_{corr}$) and Pearson ($P_{corr}$) correlation coefficients equal $0.63$ and $0.62$, respectively.} 
    \label{fig:TPR_ModularityNorm}
\end{figure*}
We see that the correlation between the rankings of the graphs obtained from the two scores is poor, except when we consider only networks produced by the LFR model or SBM.

Since the maximum GN modularity can be fairly large on random graphs~\cite{guimera04}, a potentially useful quality function could be defined by subtracting a random baseline from it. A natural option is computing the average maximum modularity over a set of randomizations of the network at hand. We use randomizations that preserve the degree sequence of the network, to be consistent with modularity's null model. The average of the maximum modularity over such graphs is $\langle Q_{rand}^{\max}\rangle$. We define the \textit{modularity difference} of the network as the difference between its maximum GN modularity and $\langle Q_{rand}^{\max}\rangle$
\begin{equation}
    Q_{diff}=Q^{\max}-\langle Q_{rand}^{\max}\rangle.
\end{equation}
\begin{figure*}[htb]
    \centering
    \includegraphics[width=0.95\textwidth]{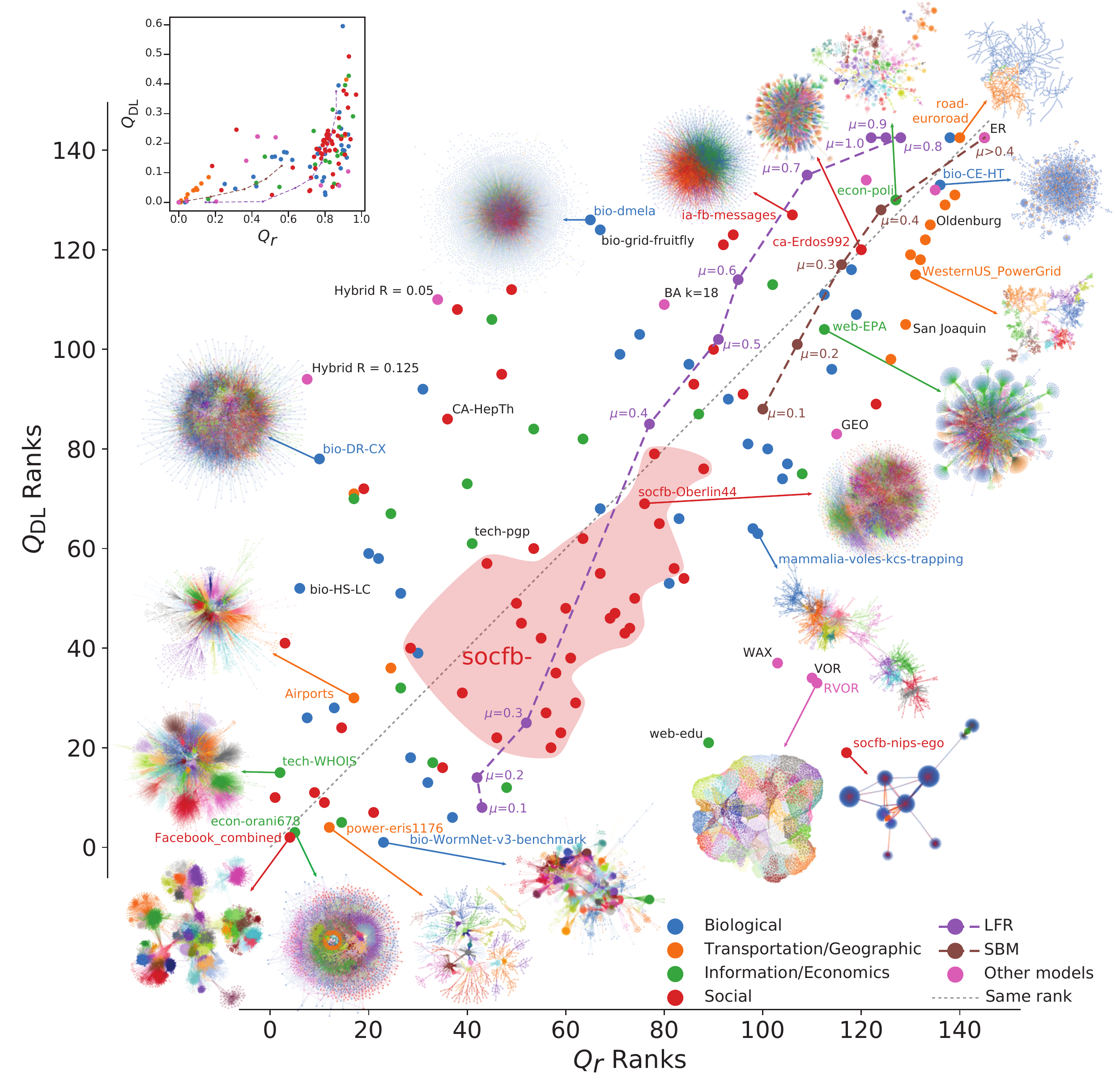}
    \caption{Relationship between $Q_r$ and $Q_{DL}$. The diagram is analogous to the one in Fig.~\ref{fig:TPR_ModularityNorm}. The Spearman ($S_{corr}$) and Pearson ($P_{corr}$) correlation coefficients are $0.69$ and $0.75$, respectively.} 
    \label{fig:TPR_MDL}
\end{figure*}

To compute $Q_{diff}$ we created 200 randomizations of the network. The modularity maximum for each randomization is approximated by the largest value of $Q$ found on $10$ partitions obtained by running the Louvain algorithm. Figure \ref{fig:TPR_ModularityNorm} shows that $Q_r$ and $Q_{diff}$ are strongly correlated. The main outliers are the 
artificial model networks consisting of a clique connected to a random graph, that we have examined in Section~\ref{sec:modnet}. In this case, unless the clique is really small compared to the graph size, $Q_r$ may take appreciable values, as it is hard to disrupt the clique via rewiring, whereas according to $Q_{{diff}}$ the network is essentially a random graph, so it has a low score.

An additional measure that can be used as a quality function is associated to the statistical inference clustering technique we have used. Let DL($P$) be the description length of the partition $P$, detected via the SBM, and of the model parameters. The description length for the trivial partition $T$ in one cluster and the model parameters is DL($T$). We define the \textit{information modularity} $Q_{{DL}}$ as

\begin{equation}
    Q_{{DL}}=1-\frac{{DL}(P)}{{DL}(T)}.
    \label{eq:QDL}
\end{equation}

\begin{figure*}[htb]
    \centering
    \includegraphics[width=0.95\textwidth]{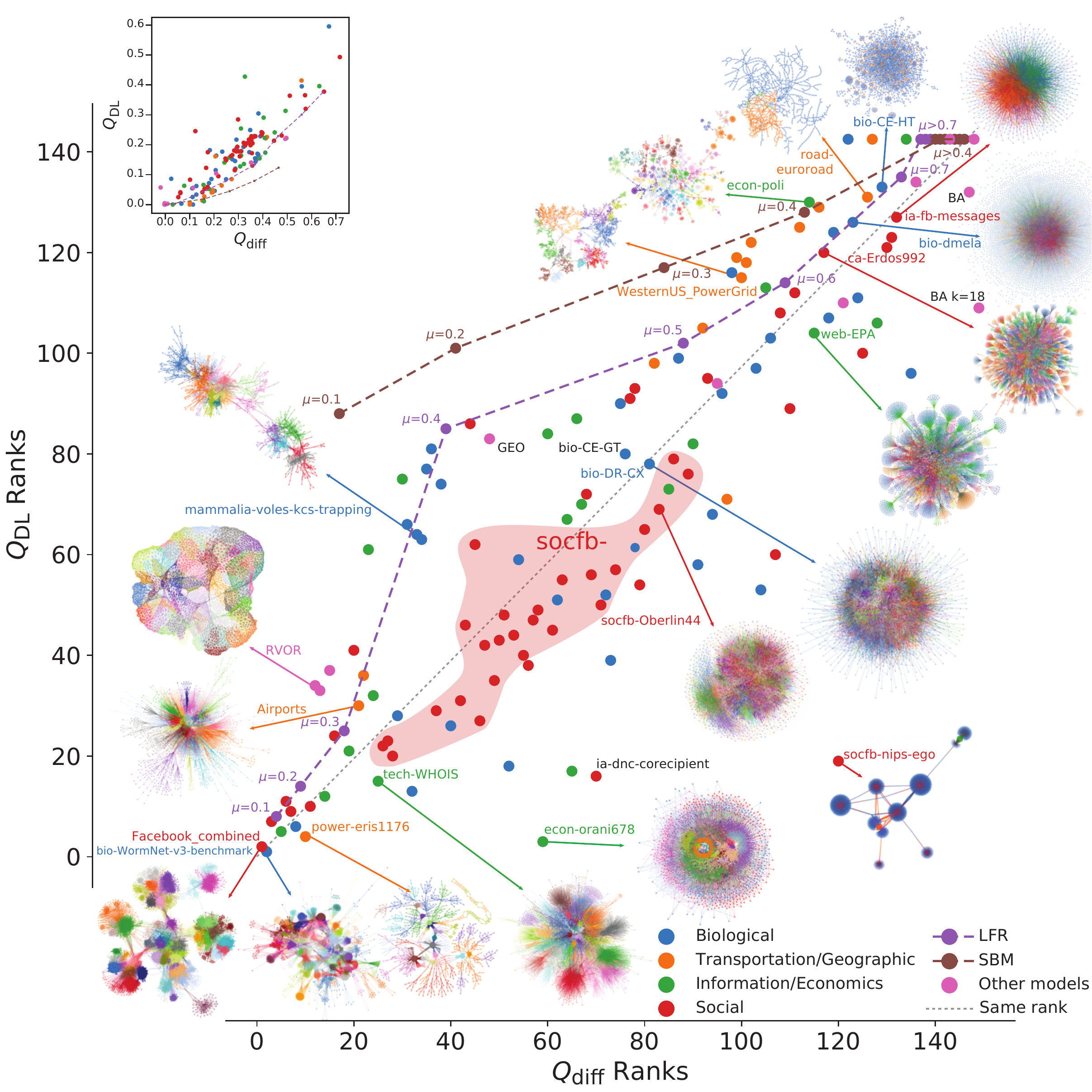}
    \caption{Relationship between $Q_{diff}$ and $Q_{DL}$.
    The diagram is analogous to the ones in Figs.~\ref{fig:TPR_ModularityNorm} and \ref{fig:TPR_MDL}.
    The Spearman ($S_{corr}$) and Pearson ($P_{corr}$) correlation coefficients are $0.88$ and $0.86$, respectively.} 
    \label{fig:MDL_ModularityNorm}
\end{figure*}

The description length of the degree-corrected microcanonical SBM is defined in Appendix~\ref{apx_MDL}.
The variable $Q_{{DL}}$ basically expresses a sort of ``distance" between $P$ and $T$ in terms of information content. When $P=T$, $Q_{{DL}}=0$. If $P$ is clearly better than $T$, $DL(P)\ll {DL}(T)$ and $Q_{{DL}}\sim 1$. We stress that this measure can be extended to any statistical inference technique based on generative network models, as long as the description length is properly defined.

Figure~\ref{fig:TPR_MDL} indicates that
$Q_{{DL}}$ is strongly correlated with $Q_r$. As in Fig.~\ref{fig:TPR_ModularityNorm} the model networks consisting of a clique connected to a random graph are major outliers, for the same reason exposed above. Unsurprisingly then, $Q_{{diff}}$ and $Q_{{DL}}$ are very highly correlated (Fig.~\ref{fig:MDL_ModularityNorm}), which sets a non-trivial link between modularity optimization and stochastic blockmodeling.

We encourage to use $Q_r$, $Q_{{diff}}$ and $Q_{{DL}}$ for the assessment of network modularity. A big advantage of $Q_{{diff}}$, $Q_{{DL}}$ over $Q_r$ is that they can be computed much faster, as they do not require the derivation of the TPR profile, which is costly.

\section{Discussion}

We have proposed a new concept to define the strength of the partition of a network in communities, that does not suffer from drawbacks of earlier proposals, most notably the modularity by Newman and Girvan. We interpret the modularity of a graph as a measure of the robustness of its communities
against perturbations of its structure. We quantify this criterion via the trivial partition ratio (TPR), indicating the probability that the graph has no community structure as a function of the extent $p$ of the perturbation, which expresses the fraction of the links that are randomly rewired. We find that most graphs lose their group structure within a narrow range of $p$. Therefore, the diagram showing the variation of the TPR with $p$ can be used to define the robustness of a graph's community structure.

Results on artificial benchmark graphs with built-in clusters agree with intuition. Our analysis of real datasets shows that, for large classes of networks, RM has a broad distribution, so we cannot claim, e.g., that social networks are more modular than biological ones, or vice versa. 

RM can be quantified by using any clustering algorithm that can recognize the absence of communities. The actual scores would depend on the specific method adopted, but the rankings between graphs should be relatively stable. In fact, since what matters for our definition is whether a clustering method finds or not communities, without considering their features, the relevant difference between methods is how prompt they are to realize that communities have been disrupted, which translates into a different threshold for the level of noise that makes the detection impossible. We do not expect that such systematic shifts in the thresholds would significantly alter the modularity rankings among different networks.
Unfortunately, we cannot support this expectation with data, because finding clustering algorithms that correctly recognize community-free networks is challenging. Popular techniques, like Louvain and Infomap, find clusters in Erd\H{o}s-R\'enyi random graphs, for instance. This is an important drawback of such methods in general and we urge the scientific community to pay more attention to this issue looking forward. In our calculations, we have used a posteriori stochastic blockmodeling, but we strongly encourage to explore other options as well. Among the other things, alternative techniques might overcome the drawbacks we have encountered with SBMs, like the inability to find disassortative structure and the detection of spurious partitions (Appendix A).

Estimating RM requires the generation of several perturbed networks, along with the detection of their partitions, which limits the size of networks that can be analyzed. On the other hand, since all randomizations and all runs of the clustering algorithms are independent, the calculation is highly parallelizable.

The RM $Q_r$ is uncorrelated with the GN modularity $Q$, reaffirming well-known concerns about the use of the latter to quantify community structure. However, we have found that the modularity difference $Q_{diff}$, obtained by subtracting a random baseline from 
$Q$, has a strong correlation with $Q_r$. Likewise, the information modularity $Q_{DL}$, expressing a sort of distance in terms of information content between the detected partition and the trivial division into a single cluster, is also strongly correlated with $Q_r$ and very strongly correlated with $Q_{diff}$, which discloses a deep connection between two apparently unrelated classes of graph clustering approaches: modularity optimization and stochastic blockmodeling.

The measures $Q_r$, $Q_{diff}$ and $Q_{DL}$ rely on different notions of modularity. The RM $Q_r$ estimates the robustness of the most pronounced cluster, whereas the other two measures give an assessment of the whole partition. Indeed, if one or a few pronounced clusters stick out of an otherwise weakly modular graph, $Q_r$ tends to be comparatively higher than the other two scores, whose assessment would put more weight on the rest of the graph, which is much larger than the clusters. Still, the high correlations we found between $Q_r$, on the one side, and $Q_{diff}$ and $Q_{DL}$, on the other, show that in real networks such scenario is not frequent. Hence, the distinct concepts of modularity yield fairly similar rankings of networks. In general, we recommend to explore the potential of $Q_{diff}$ and $Q_{DL}$ as well. The code to calculate $Q_r$, $Q_{diff}$ and $Q_{DL}$ for any input network is freely available here: \href{https://github.com/filipinascimento/RModularity}{github.com/filipinascimento/RModularity}.

\section*{Acknowledgments}

This project was funded by the Deanship of Scientific Research (DSR) at King Abdulaziz University, Jeddah, Saudi Arabia, under Grant No. RG-1439-311-10. AA, VT, WA and SF therefore, acknowledge with thanks DSR for technical and financial support. In addition, this project was partially supported by the Army Research Office under contract number W911NF-21-1-0194 and by the Air Force Office of Scientific Research under award number FA9550-19-1-0391. Research was carried out using computational resources of the Indiana University Network Science Institute (IUNI).

% \section{Appendix}
\appendix
%[????]
% STILL TO DO HERE:
% \begin{itemize}
    
%     % \item We also considered network models without communities, they are in the SI. Separated panels. F I THINK ONE PANEL IS OK, BUT WE NEED TO DESCRIBE BRIEFLY THE MODELS IN THE SI (ONE SHORT PARAGRAPH FOR EACH MODEL)
%     % \item Include formulas for DL (copy from Thiago's arxiv.)
    
% \end{itemize}

\section{Synthetic Networks Models}
\renewcommand{\thefigure}{A\arabic{figure}}
\setcounter{figure}{0}
\renewcommand{\thetable}{A\arabic{table}}
\setcounter{table}{0}
\label{apx_models}

\subsection{The LFR model}
\label{subsec:LFR}

The LFR benchmark~\cite{lancichinetti08} is characterized by power-law distributions of degree and community size, reflecting the heterogeneity of these two variables in real graphs. 
The parameters needed to generate LFR graphs are the number of nodes $N$, the exponents of the distributions of degree ($\tau_1$) and community size ($\tau_2$), the average degree $\langle k\rangle$, the maximum degree $k_{max}$, the extremes of the range of community sizes $c_{min}$ (lower) and $c_{max}$ (upper), and the mixing parameter $\mu$, that we defined in Section~\ref{sec:modnet}.

For our tests we used the following set of parameters: $N = 5,000$,
$\tau_1 = 2$,
$\tau_2 = 1$,
$\langle k\rangle = 10$,
$k_{max} = 250$,
$c_{min} = 20$,
$c_{max} = 500$.

\begin{figure}[htb]
    \centering
    \includegraphics[width=\columnwidth]{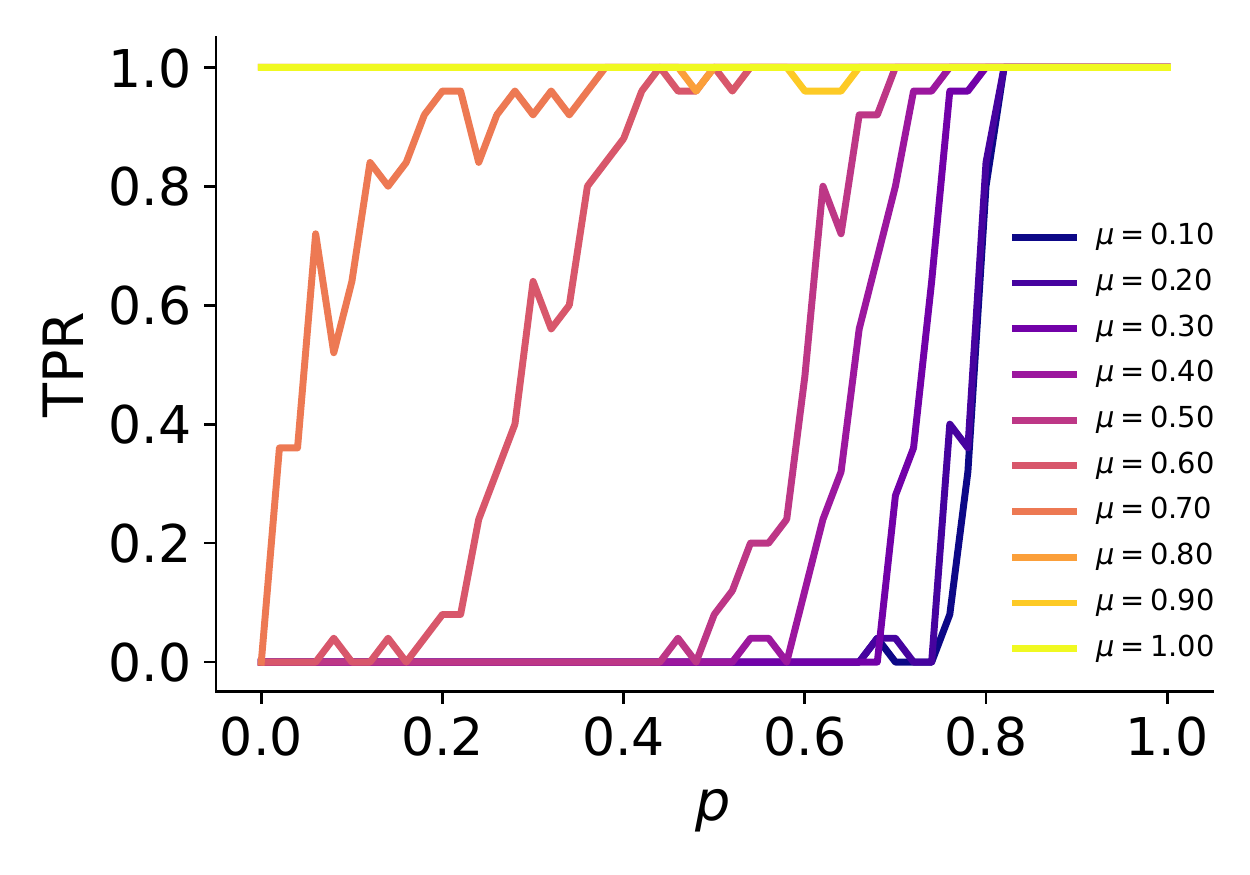}
    \caption{Trivial partition ratio (TPR) as a function of $p$ for LFR benchmark graphs for different values of the mixing parameter $\mu$.} 
    \label{fig:TPR_LFR_Models}
\end{figure}

In Fig.~\ref{fig:TPR_LFR_Models} we show the TPR profiles for different values of $\mu$. The behavior is qualitatively the same as for the SBM benchmark graphs (Fig.~\ref{fig:TPR_Benchmarks}). 
Like for the SBM benchmarks, the SBM-based clustering algorithm we have used is not able to detect the planted disassortative partition in the unperturbed graphs ($p=0$) with large $\mu$, not even when $\mu=1$. We will look into that in Appendix C. 

As we experimented with different parameter sets, specifically for larger values of the average degree $\langle k\rangle$, we occasionally found spurious core-periphery structures for large $\mu$-values, causing a dip of the TPR in the central part of the diagram, followed by a return to the plateau. Such core-periphery structure emerges from the fact that, when the LFR graph is highly disassortative, hubs are strongly connected to each other. As the graph is randomized, such hub structure is harder to destroy than the organization of the rest of the network, so it may emerge as a group, with the other nodes being put in the other group. Eventually, the randomization ends up destroying this hub structure as well, so, for sufficiently large $p$, the algorithm returns the trivial division again. Still, from the point of view of the DL, we found that the qualities of the core-periphery structure and the trivial partition are basically the same. Such anomalous behavior occurs in none of the TPR profiles of the real datasets.

\subsection{Other random network models}

On top of LFR and SBM benchmark networks we also employed the following random network models:

{\bf Erd{\H o}s-Rényi model (ER)}~\cite{erdos59}: Homogeneous network model with $N$ nodes. Pairs of nodes are connected according to a probability $p$, leading to $E = p N(N-1)/2$ links for undirected networks. ER can be understood as the simplest random network model assuming a given number of nodes and average degree.

{\bf Barabási–Albert model (BA)}~\cite{barabasi99}: Networks are generated through two concurrent mechanisms: growth and preferential attachment. The network starts from a fully connected graph of $m_{initial}$ nodes. For each iteration, a new node is added to the network and connected to $m$ of the existing nodes, chosen with a probability proportional to their current degrees. The process is repeated $n$ ($>> m_{initial}$) times leading to $N=m_{initial}+n$ nodes and $E=m_{initial}(m_{initial}-1)/2 + m*n$ links. This model reproduces the power-law degree distribution found in many real-world systems. 

{\bf Voronoi model (VOR)}~\cite{barthelemy2011spatial}: A Voronoi network is created by placing $N$ nodes randomly over a 2D space and by applying the Delaunay triangulation algorithm to find the links. This type of network by construction is planar but not regular and represents the cells and neighbors in a Voronoi diagram.

{\bf Rewired Voronoi model (RVOR)}: First, a VOR network is created. Next, the links of the network are randomly rewired according to a probability $p_{rewire}$. This procedure is similar to generating a Watts-Strogatz network~\cite{watts1998collective}, however with both ending points of the rewired links being uniformly chosen among all the nodes in the network. Here we adopted $p_{rewire} = 0.01$, which can generate networks with a strong local structure, as well as presenting shortcuts.

{\bf Geographic model (GEO)}~\cite{barthelemy2011spatial}: Can be generated by placing $N$ nodes across a 2D space and connecting all the pairs that are at most within a distance $d_{threshold}$ from each other. This model generates networks with strong local structure, but no possibility for shortcuts.

{\bf Waxman model (WAX)}~\cite{waxman1988routing,barthelemy2011spatial}: WAX networks are generated in a similar fashion as the GEO model, but instead of connecting nodes within a fixed threshold distance, pairs of nodes $i,j$ are connected according to a probability $\pi_{ij}=p \exp{-(d_{ij}/\beta)}$ where $\beta$ and $p$ are parameters of the model and $d_{ij}$ is the distance between $i$ and $j$. Here we adopted $\beta = 1.0$ and $p = 0.05$.

\begin{figure}[htbp]
    \centering
    \includegraphics[width=\columnwidth]{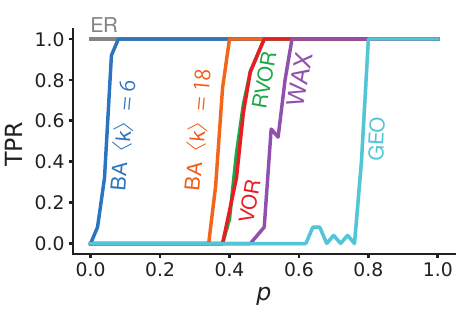}
    \caption{TPR curves for various random network models.} %will change Girvan-Newman to SBM
    \label{fig:TPR_Models}
\end{figure}

The TPR curves obtained for the models above are shown in Figure~\ref{fig:TPR_Models}. ER graphs are not modular, as expected, so we get a flat profile. VOR, RVOR, GEO and WAX, despite the randomness underlying their construction processes, display a kind of modular structure, because links are local, as in lattices, and we know of no clustering algorithm which does not break a lattice. Surprisingly, the profile for the BA models, which should generate networks without communities, is not flat: the plateau is reached around a threshold value which shifts to the right as the average degree $\langle k\rangle$ increases. Our SBM-based clustering algorithm finds two groups for $p=0$, denoting a core-periphery structure. By construction, since it is a growing network model, the oldest nodes are also the ones with the largest degree, hence hubs are strongly connected with each other, forming a structure that misleads the algorithm. This problem is related to the one that occasionally occurs for LFR benchmark graphs as well (Section~\ref{subsec:LFR}), for special parameter choices.

\section{Description length}
\label{apx_MDL}

\renewcommand{\thefigure}{B\arabic{figure}}
\setcounter{figure}{0}
\renewcommand{\thetable}{B\arabic{table}}
\setcounter{table}{0}

We calculate $Q_{{DL}}$ based on the description length (${DL}$) defined in \cite{peixoto13}. The description length (DL) can be defined as the entropy of the microcanonical SBM ensemble $\mathcal{S}$ of the graphs generated by the model, plus the entropy $\mathcal{L}$ related to the information needed to describe the model itself
\begin{equation}\label{eq:DL}
  {DL} = \mathcal{S} + \mathcal{L}.
\end{equation}
For the microcanonical degree-corrected SBM that we have used these quantities are approximated by the following expressions:
\begin{equation}\label{eq:degree}
  \mathcal{S} \cong -E -\sum_kN_k\ln k! - \frac{1}{2} \sum_{rs}e_{rs}\ln\left(\frac{e_{rs}}{e_re_s}\right),
\end{equation}
\begin{equation}\label{eq:lt}
  \mathcal{L} \cong Eh\left[\frac{q(q+1)}{2E}\right] + N\ln q - N\sum_kp_k\ln p_k,
\end{equation}
where $N$ and $E$ are respectively the number of nodes and links of the network, $n_r$ the number of nodes in a group $r$, $N_k$ is the number of nodes with degree $k$, $e_{rs}$ is the block matrix, i.e., the number of links between nodes of groups $r$ and $s$, $e_r=\sum_se_{rs}$, $h[x] = (1+x)\ln (1+x) -x\ln x $, $q$ is the number of groups, and $p_k$ is the proportion of nodes with degree $k$. The DL of the trivial partition into one cluster is obtained by setting $q=1$. 

\section{Disassortative communities}
\label{apx_disassortative}
\renewcommand{\thefigure}{C\arabic{figure}}
\setcounter{figure}{0}
\renewcommand{\thetable}{C\arabic{table}}
\setcounter{table}{0}

In Figs.~\ref{fig:TPR_Benchmarks} and \ref{fig:TPR_LFR_Models} we have seen that our SBM-based clustering algorithm did not detect disassortative communities. For $\mu=1$ the task becomes identical to the \textit{graph coloring problem}, i.e. the problem of assigning colors to nodes, such that each node has different colors than its neighbors~\cite{jensen11}. This task is NP-hard, so it is not surprising that algorithms may fail at it. Still, since the division found by our SBM algorithm is obtained by using Markov Chain Monte Carlo~\cite{newman99}, it is possible, e.g. that the algorithm stops before equilibrium is reached, so that the division it finds is suboptimal. To check for that, we need to compare the goodness of the trivial partition $T$ in one cluster, the planted partition $P$ and the one detected by the algorithm ($D$). This can be done by comparing the values of their description length (DL), which we computed via the function \textit{entropy()} of Graph-tool v2.35~\cite{peixoto_graph-tool_2014}
%(Appendix B, Eqs.~\ref{eq:DL}, \ref{eq:degree} and \ref{eq:lt}).

\begin{figure}[htb]
    \centering
    \includegraphics[width=\columnwidth]{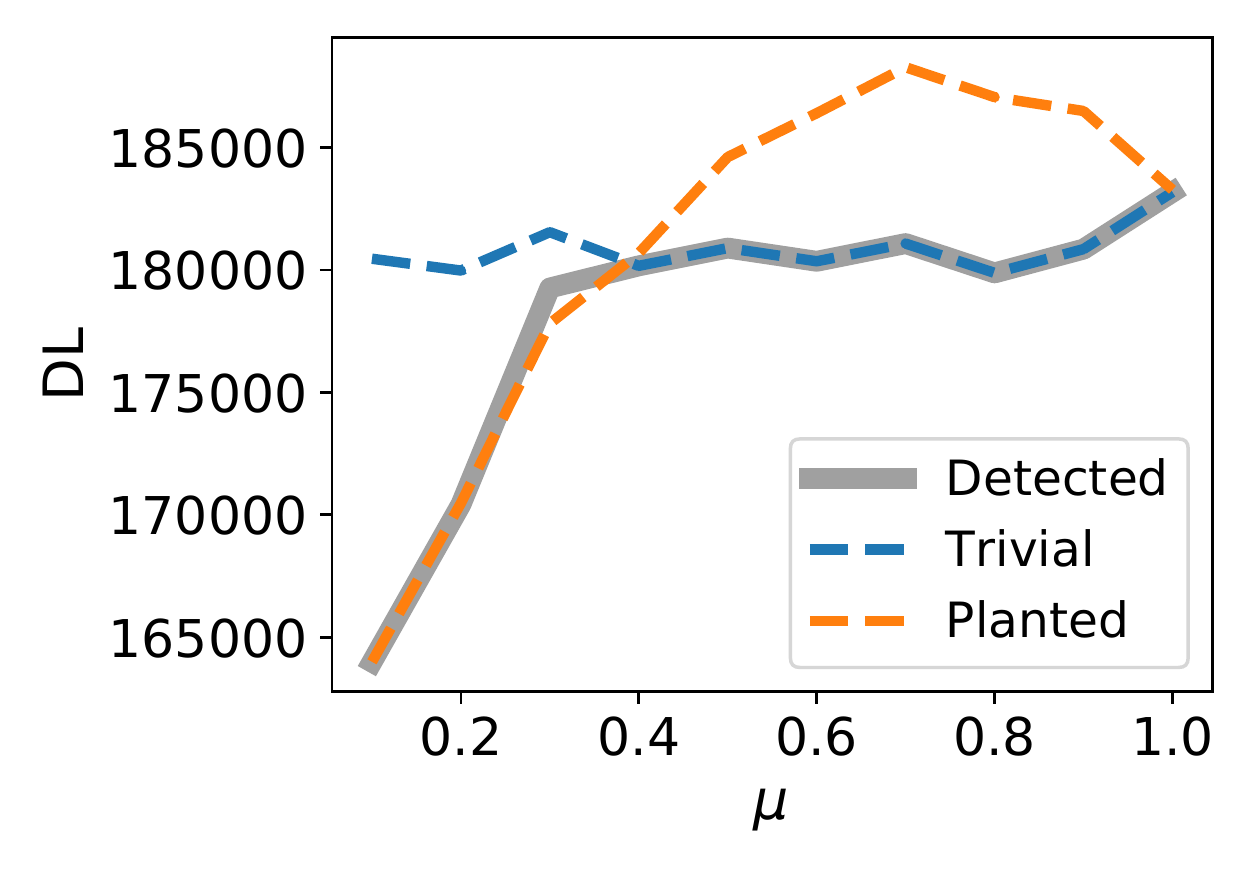}
    \caption{Description lengths for three divisions obtained on unperturbed SBM benchmark networks ($p=0$), as a function of the mixing parameter $\mu$: the partition detected by the SBM clustering method, the planted partition and the trivial partition.}
    \label{fig:MDL3GN}
\end{figure}
\begin{figure}[htb]
    \centering
    \includegraphics[width=\columnwidth]{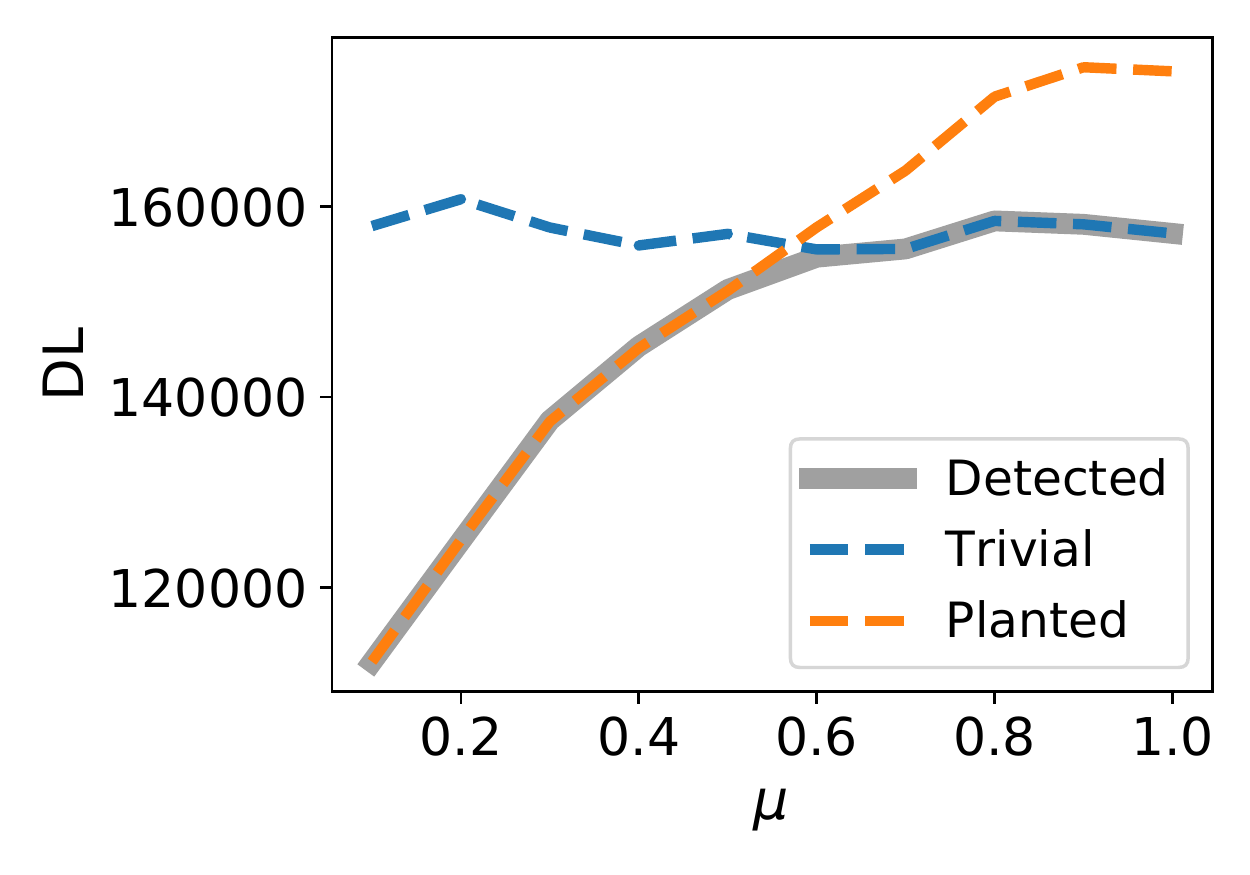}
    \caption{Description lengths for three divisions obtained on unperturbed LFR benchmark networks ($p=0$), as a function of the mixing parameter $\mu$: the partition detected by the SBM clustering method, the planted partition and the trivial partition.}
    \label{fig:MDL3}
\end{figure}

In Fig.~\ref{fig:MDL3GN} we show the DL of the three divisions as a function of $\mu$ for the SBM benchmarks ($p=0$). We find two relevant ranges of $\mu$. In the first range, going from $\mu=0$ to $\mu \approx 0.4$, the algorithm finds indeed the planted (assortative) partition of the graph, up to a small fluctuation around $\mu=0.3$. 
%For $\mu$ between $\sim 0.5$ and $\sim 0.9$, $T$ has the shorter $DL$ and is correctly identified by the algorithm, so here there are no communities, neither assortative nor disassortative, as expected. 
In the second range, for $\mu$ between $\sim 0.4$ and $1$, $P$ has the higher $DL$, so it is a worse solution than $T$. So, in this case, the method is not overlooking $P$, but it finds the better solution $T$. Still, we can see that, for $\mu=1$, the DL of the planted disassortative partition is extremely close to the DL of $T$. In fact, we have found that, for $\langle k\rangle >10$, the disassortative partition is the better solution, at least for $\mu=1$, and that the method is unable to find it.% possibly due to convergence issues of the Markov chain Monte Carlo (MCMC).
%or to an overwhelming proportion of (slightly) suboptimal solutions, which are then much more likely to be found.

%\begin{figure}[htb]
%    \centering
%    \includegraphics[width=\columnwidth]{Figures/MDLDiff_byP.pdf}
%    \caption{Average information modularity ($Q_{DL}$) for an LFR network with $\mu=1$, as a function of the rewiring probability $p$.}
%    \label{fig:MDL4}
%\end{figure}

In Fig.~\ref{fig:MDL3} we show the same plot for LFR benchmark graphs. We observe a similar pattern as in Fig.~\ref{fig:MDL3GN}, with two relevant ranges of $\mu$.
%Specifically, for large $\mu$-values, the planted disassortative partition has higher DL than $T$, which is the one found by the algorithm. Hence, we do detect the better division. 
However, here there is a marked gap between the DL of the planted disassortative partition and the DL of $T$, over the whole range. In addition, we have found that this gap does not rapidly decrease as we increase the average degree, as it happens instead for the SBM graphs. Why the planted disassortative partition is suboptimal for the LFR graphs deserves further investigation.

%Finally, we use the same approach to investigate the anomalies in the TPR profiles of LFR benchmarks for high $\mu$-values (Fig.~\ref{fig:TPR_LFR_Models}). Now we show the difference between the DL of $T$ and $D$, divided by the DL of $T$, as a function of $p$ for a fixed value of the mixing parameter, $\mu=1$ (Fig.~\ref{fig:MDL4}). We remark that this is just the information modularity $Q_{DL}$ (Eq.~\ref{eq:QDL}). As we can see, for low and high values of $p$, $D=T$, so $Q_{DL}$ is zero. Instead, in the middle range of $\mu$, where the TPR curve displays a drop, $D\neq T$, and reveals a core-periphery structure. Such division has lower DL than $T$, albeit the relative difference in the DLs is really tiny. Hence, the division found by the algorithm is indeed better than $T$.  The spurious core-periphery structure emerges from the fact that, when the LFR graph are highly disassortative, hubs are strongly connected to each other. As the graph is randomized, such hub structure is harder to destroy than the organization of the rest of the network, so it emerges as a group, while the other nodes are put in the other group. Eventually the randomization ends up destroying this hub structure as well, so, for sufficiently large $p$, the algorithm returns the trivial division again. Still, from the point of view of the DL, the qualities of the core-periphery structure and the trivial partition are basically the same.

\section{TPR profiles of real networks}
\label{apx_TPR_ALL}
\renewcommand{\thefigure}{D\arabic{figure}}
\setcounter{figure}{0}
\renewcommand{\thetable}{D\arabic{table}}
\setcounter{table}{0}
Here we show the TPR curves for all real network datasets we analyzed, divided by category (Figs.~\ref{fig:TPR_All_Biological}-\ref{fig:TPR_All_Information_Economy}). We also rank the networks according to the values of $Q_r$ in Tables~\ref{table:biological}-\ref{table:information_economy}.

\begin{figure*}[htb]
    \centering
    \includegraphics[width=0.95\textwidth]{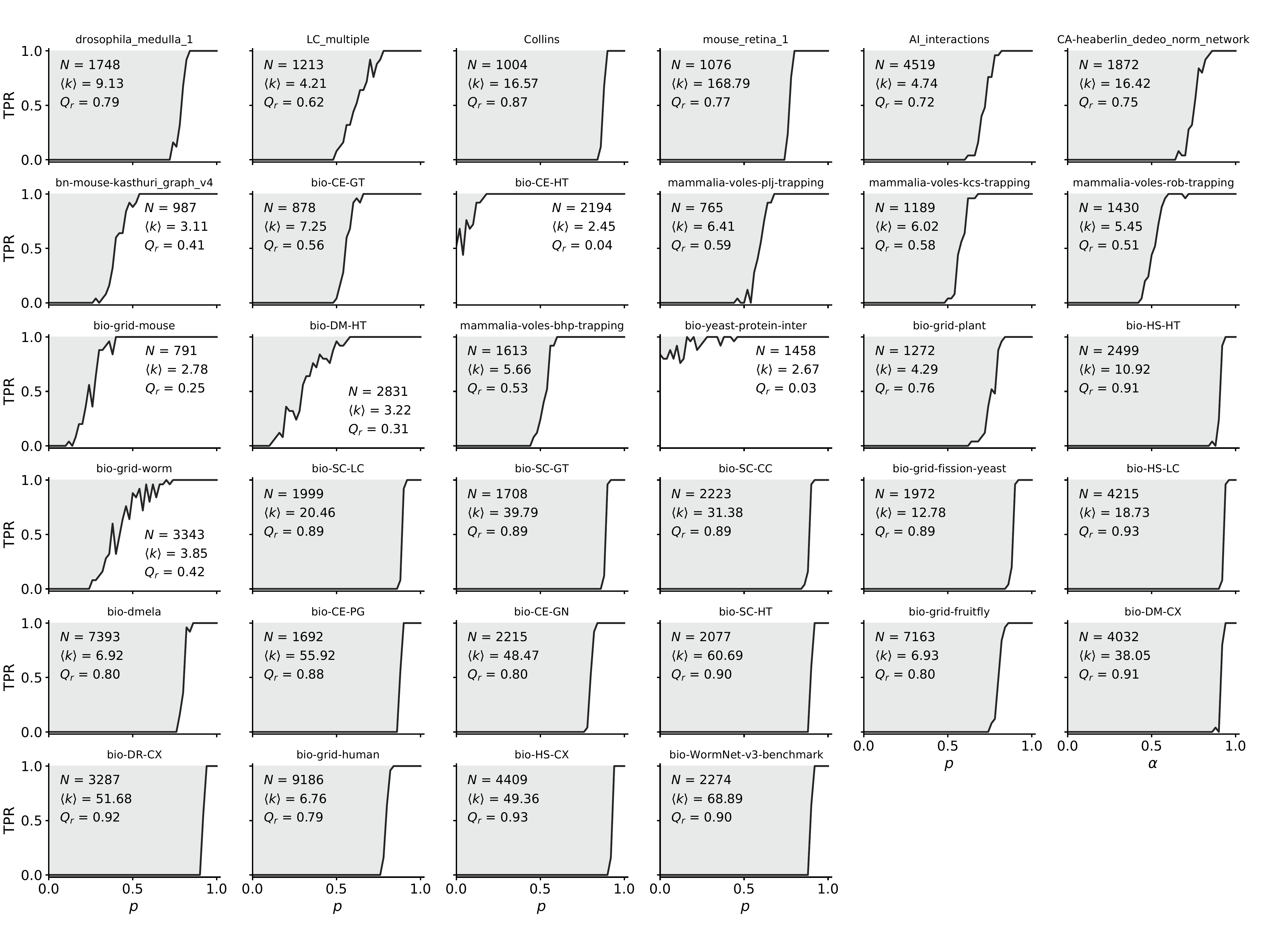}
    \caption{TPR curves for biological networks.}
    \label{fig:TPR_All_Biological}
\end{figure*}

\begin{figure*}[htb]
    \centering
    \includegraphics[width=0.95\textwidth]{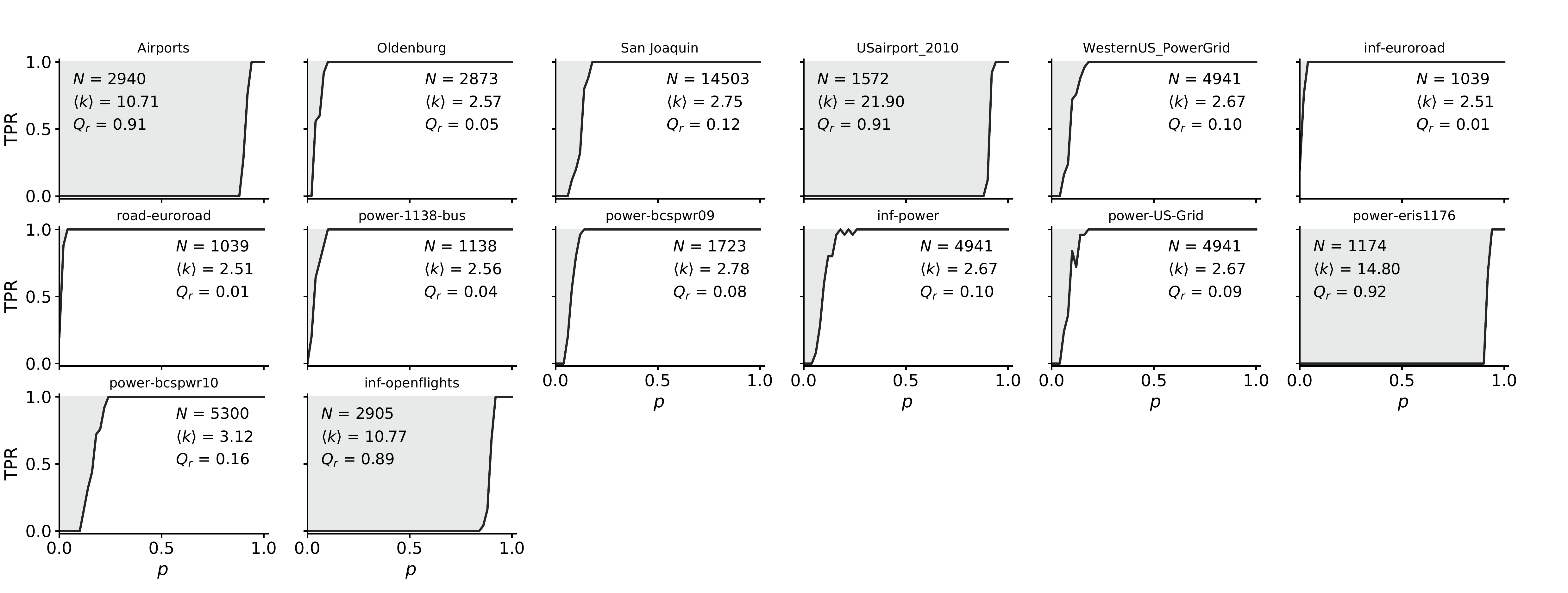}
    \caption{TPR curves for the transportation/geographic networks.}
    \label{fig:TPR_All_Transportation_Geographic}
\end{figure*}

\begin{figure*}[htb]
    \centering
    \includegraphics[width=0.95\textwidth]{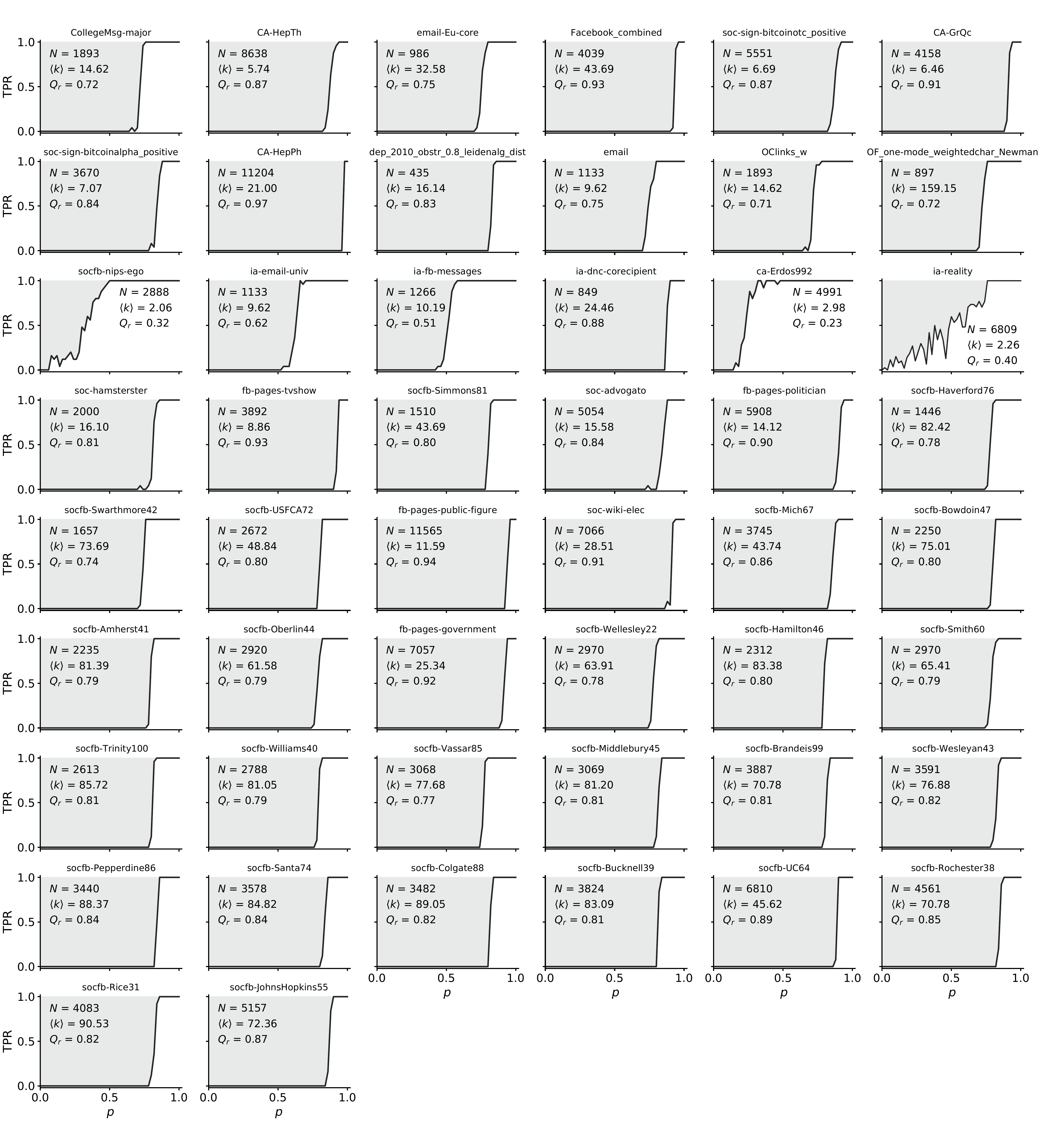}
    \caption{TPR curves for social networks.}
    \label{fig:TPR_All_Social}
\end{figure*}

\begin{figure*}[htb]
    \centering
    \includegraphics[width=0.95\textwidth]{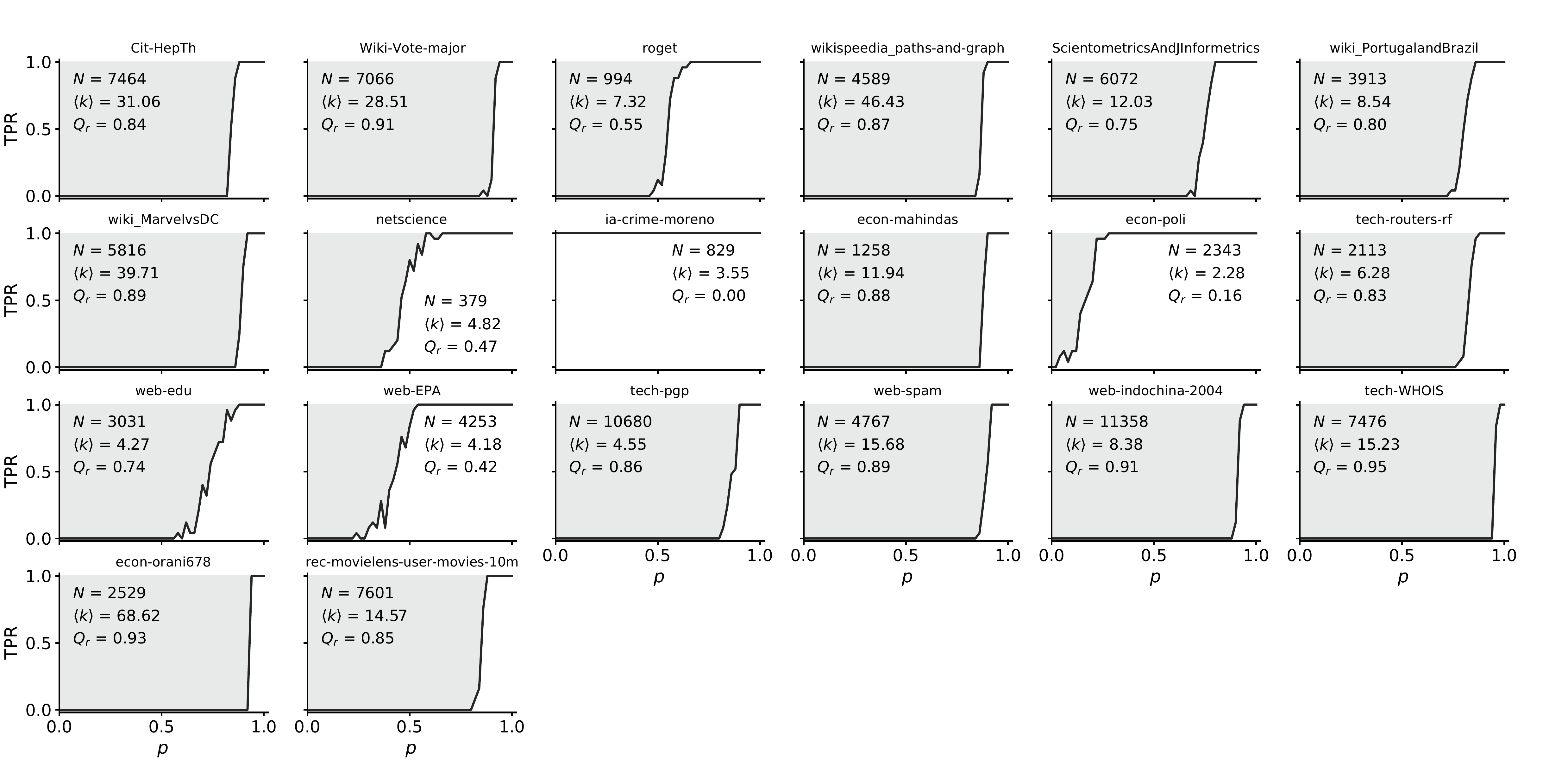}
    \caption{TPR curves for information/economic networks.}
    \label{fig:TPR_All_Information_Economy}
\end{figure*}

% \begin{figure}[!h]
%     \centering
%     \includegraphics[width=0.8\linewidth]{Figures/TPR_byCategory_basic_LFR.pdf}
%     \caption{TPR curves for the LFR networks.}
%     \label{fig:TPR_All_LFR}
% \end{figure}
% \begin{figure}[!h]
%     \centering
%     \includegraphics[width=0.8\linewidth]{Figures/TPR_byCategory_basic_SBM.pdf}
%     \caption{TPR curves for the SBM networks.}
%     \label{fig:TPR_All_SBM}
% \end{figure}
% \begin{figure}[!h]
%     \centering
%     \includegraphics[width=0.8\linewidth]{Figures/TPR_byCategory_basic_Models.pdf}
%     \caption{TPR curves for the Model networks.}
%     \label{fig:TPR_All_Models}
% \end{figure}

% TPR_byCategory_basic_LFR.pdf
% TPR_byCategory_basic_Information_Economy.pdf
% TPR_byCategory_basic_Transport-Geographic.pdf
% TPR_byCategory_basic_Social.pdf
% TPR_byCategory_basic_Models.pdf
% TPR_byCategory_basic_SBM.pdf
% TPR_byCategory_basic_Biological.pdf

% \begin{itemize}

%     % \item Model networks without communities (BA, ER, Hybrid, Grid-like). 
    
%     % \item Degree preserving randomization with the same plots of the 
    
%     % \item Hybrid model TPR, MDL, etc.
    
% \end{itemize}

\newpage 

\begin{table*}[htb]
\begin{center}
%\centering
\caption{Ranking of biological networks based on their values of $Q_r$.}
\footnotesize
\label{table:biological}
\input{Tables/table_TPR_byCategory_basic_Biological.tex}
\end{center}
\end{table*}

\begin{table*}[htb]
\centering
\caption{Ranking of transportation and geographic networks based on their values of $Q_r$.}
\footnotesize
\label{table:transportation_geographic}
\input{Tables/table_TPR_byCategory_basic_Transport-Geographic}
\end{table*}

\begin{table*}[htb]
\centering
\caption{Ranking of social networks based on their values of $Q_r$.}
\footnotesize
\label{table:social}
\input{Tables/table_TPR_byCategory_basic_Social}
\end{table*}

\begin{table*}[htb]
\centering
\caption{Ranking of information and economic networks based on their values of $Q_r$.}
\footnotesize
\label{table:information_economy}
\input{Tables/table_TPR_byCategory_basic_Information_Economy}
\end{table*}

\bibliography{references}

\end{document}

%% file: Tables/table_TPR_byCategory_basic_Biological.tex
\renewcommand{\arraystretch}{0.5}
\setlength{\tabcolsep}{2em}
\begin{tabular}{r c c c}
\toprule
                         Network &    $N$ & $\langle k \rangle$ & $Q_r$ \\
\hline
                       bio-HS-LC & 4215 &               18.7 &  0.93 \\
                       bio-HS-CX & 4409 &               49.4 &  0.93 \\
                       bio-DR-CX & 3287 &               51.7 &  0.92 \\
                       bio-DM-CX & 4032 &               38.1 &  0.91 \\
                       bio-HS-HT & 2499 &               10.9 &  0.91 \\
                       bio-SC-HT & 2077 &               60.7 &  0.90 \\
        bio-WormNet-v3-benchmark & 2274 &               68.9 &  0.90 \\
                       bio-SC-LC & 1999 &               20.5 &  0.89 \\
                       bio-SC-GT & 1708 &               39.8 &  0.89 \\
                       bio-SC-CC & 2223 &               31.4 &  0.89 \\
          bio-grid-fission-yeast & 1972 &               12.8 &  0.89 \\
                       bio-CE-PG & 1692 &               55.9 &  0.88 \\
                         Collins & 1004 &               16.6 &  0.87 \\
                       bio-dmela & 7393 &                6.9 &  0.80 \\
               bio-grid-fruitfly & 7163 &                6.9 &  0.80 \\
                       bio-CE-GN & 2215 &               48.5 &  0.80 \\
                  bio-grid-human & 9186 &                6.8 &  0.79 \\
            drosophila\_medulla\_1 & 1748 &                9.1 &  0.79 \\
                  mouse\_retina\_1 & 1076 &              168.8 &  0.77 \\
                  bio-grid-plant & 1272 &                4.3 &  0.76 \\
 CA-heaberlin\_dedeo\_norm\_network & 1872 &               16.4 &  0.75 \\
                 AI\_interactions & 4519 &                4.7 &  0.72 \\
                     LC\_multiple & 1213 &                4.2 &  0.62 \\
     mammalia-voles-plj-trapping &  765 &                6.4 &  0.59 \\
     mammalia-voles-kcs-trapping & 1189 &                6.0 &  0.58 \\
                       bio-CE-GT &  878 &                7.2 &  0.56 \\
     mammalia-voles-bhp-trapping & 1613 &                5.7 &  0.53 \\
     mammalia-voles-rob-trapping & 1430 &                5.5 &  0.51 \\
                   bio-grid-worm & 3343 &                3.9 &  0.42 \\
      bn-mouse-kasthuri\_graph\_v4 &  987 &                3.1 &  0.41 \\
                       bio-DM-HT & 2831 &                3.2 &  0.31 \\
                  bio-grid-mouse &  791 &                2.8 &  0.25 \\
                       bio-CE-HT & 2194 &                2.5 &  0.04 \\
         bio-yeast-protein-inter & 1458 &                2.7 &  0.03 \\
\bottomrule
\end{tabular}

%% file: Tables/table_TPR_byCategory_basic_Transport-Geographic.tex
\renewcommand{\arraystretch}{0.5}
\setlength{\tabcolsep}{2em}
\begin{tabular}{r c c c}
\toprule
                         Network &    $N$ & $\langle k \rangle$ & $Q_r$ \\
\hline
      power-eris1176 &  1174 &               14.8 &  0.92 \\
            Airports &  2940 &               10.7 &  0.91 \\
      USairport\_2010 &  1572 &               21.9 &  0.91 \\
     inf-openflights &  2905 &               10.8 &  0.89 \\
      power-bcspwr10 &  5300 &                3.1 &  0.16 \\
         San Joaquin & 14503 &                2.8 &  0.12 \\
           inf-power &  4941 &                2.7 &  0.10 \\
 WesternUS\_PowerGrid &  4941 &                2.7 &  0.10 \\
       power-US-Grid &  4941 &                2.7 &  0.09 \\
      power-bcspwr09 &  1723 &                2.8 &  0.08 \\
           Oldenburg &  2873 &                2.6 &  0.05 \\
      power-1138-bus &  1138 &                2.6 &  0.04 \\
        inf-euroroad &  1039 &                2.5 &  0.01 \\
       road-euroroad &  1039 &                2.5 &  0.01 \\
\bottomrule
\end{tabular}

%% file: Tables/table_TPR_byCategory_basic_Social.tex
\renewcommand{\arraystretch}{0.5}
\setlength{\tabcolsep}{2em}
\begin{tabular}{r c c c}
\toprule
                         Network &    $N$ & $\langle k \rangle$ & $Q_r$ \\
\hline
                          CA-HepPh & 11204 &               21.0 &  0.97 \\
            fb-pages-public-figure & 11565 &               11.6 &  0.94 \\
                 Facebook\_combined &  4039 &               43.7 &  0.93 \\
                   fb-pages-tvshow &  3892 &                8.9 &  0.93 \\
               fb-pages-government &  7057 &               25.3 &  0.92 \\
                           CA-GrQc &  4158 &                6.5 &  0.91 \\
                     soc-wiki-elec &  7066 &               28.5 &  0.91 \\
               fb-pages-politician &  5908 &               14.1 &  0.90 \\
                        socfb-UC64 &  6810 &               45.6 &  0.89 \\
                ia-dnc-corecipient &   849 &               24.5 &  0.88 \\
                          CA-HepTh &  8638 &                5.7 &  0.87 \\
      soc-sign-bitcoinotc\_positive &  5551 &                6.7 &  0.87 \\
              socfb-JohnsHopkins55 &  5157 &               72.4 &  0.87 \\
                      socfb-Mich67 &  3745 &               43.7 &  0.86 \\
                 socfb-Rochester38 &  4561 &               70.8 &  0.85 \\
                      soc-advogato &  5054 &               15.6 &  0.84 \\
    soc-sign-bitcoinalpha\_positive &  3670 &                7.1 &  0.84 \\
                socfb-Pepperdine86 &  3440 &               88.4 &  0.84 \\
                     socfb-Santa74 &  3578 &               84.8 &  0.84 \\
 dep\_2010\_obstr\_0.8\_leidenalg\_dist &   435 &               16.1 &  0.83 \\
                  socfb-Wesleyan43 &  3591 &               76.9 &  0.82 \\
                      socfb-Rice31 &  4083 &               90.5 &  0.82 \\
                   socfb-Colgate88 &  3482 &               89.1 &  0.82 \\
                socfb-Middlebury45 &  3069 &               81.2 &  0.81 \\
                  socfb-Bucknell39 &  3824 &               83.1 &  0.81 \\
                  socfb-Brandeis99 &  3887 &               70.8 &  0.81 \\
                   soc-hamsterster &  2000 &               16.1 &  0.81 \\
                  socfb-Trinity100 &  2613 &               85.7 &  0.81 \\
                   socfb-Simmons81 &  1510 &               43.7 &  0.80 \\
                     socfb-USFCA72 &  2672 &               48.8 &  0.80 \\
                   socfb-Bowdoin47 &  2250 &               75.0 &  0.80 \\
                  socfb-Hamilton46 &  2312 &               83.4 &  0.80 \\
                   socfb-Amherst41 &  2235 &               81.4 &  0.79 \\
                  socfb-Williams40 &  2788 &               81.1 &  0.79 \\
                     socfb-Smith60 &  2970 &               65.4 &  0.79 \\
                   socfb-Oberlin44 &  2920 &               61.6 &  0.79 \\
                 socfb-Haverford76 &  1446 &               82.4 &  0.78 \\
                 socfb-Wellesley22 &  2970 &               63.9 &  0.78 \\
                    socfb-Vassar85 &  3068 &               77.7 &  0.77 \\
                     email-Eu-core &   986 &               32.6 &  0.75 \\
                             email &  1133 &                9.6 &  0.75 \\
                socfb-Swarthmore42 &  1657 &               73.7 &  0.74 \\
   OF\_one-mode\_weightedchar\_Newman &   897 &              159.2 &  0.72 \\
                  CollegeMsg-major &  1893 &               14.6 &  0.72 \\
                         OClinks\_w &  1893 &               14.6 &  0.71 \\
                     ia-email-univ &  1133 &                9.6 &  0.62 \\
                    ia-fb-messages &  1266 &               10.2 &  0.51 \\
                    ia-reality &  6809 &                2.3 &  0.40 \\
                    socfb-nips-ego &  2888 &                2.1 &  0.32 \\
                       ca-Erdos992 &  4991 &                3.0 &  0.23 \\

\bottomrule
\end{tabular}

%% file: Tables/table_TPR_byCategory_basic_Information_Economy.tex
\renewcommand{\arraystretch}{0.5}
\setlength{\tabcolsep}{2em}
\begin{tabular}{r c c c}
\toprule
                         Network &    $N$ & $\langle k \rangle$ & $Q_r$ \\
\hline
                     tech-WHOIS &  7476 &               15.2 &  0.95 \\
                  econ-orani678 &  2529 &               68.6 &  0.93 \\
             web-indochina-2004 & 11358 &                8.4 &  0.91 \\
                Wiki-Vote-major &  7066 &               28.5 &  0.91 \\
                       web-spam &  4767 &               15.7 &  0.89 \\
                wiki\_MarvelvsDC &  5816 &               39.7 &  0.89 \\
                  econ-mahindas &  1258 &               11.9 &  0.88 \\
    wikispeedia\_paths-and-graph &  4589 &               46.4 &  0.87 \\
                       tech-pgp & 10680 &                4.6 &  0.86 \\
  rec-movielens-user-movies-10m &  7601 &               14.6 &  0.85 \\
                      Cit-HepTh &  7464 &               31.1 &  0.84 \\
                tech-routers-rf &  2113 &                6.3 &  0.83 \\
         wiki\_PortugalandBrazil &  3913 &                8.5 &  0.80 \\
 ScientometricsAndJInformetrics &  6072 &               12.0 &  0.75 \\
                        web-edu &  3031 &                4.3 &  0.74 \\
                          roget &   994 &                7.3 &  0.55 \\
                     netscience &   379 &                4.8 &  0.47 \\
                        web-EPA &  4253 &                4.2 &  0.42 \\
                      econ-poli &  2343 &                2.3 &  0.16 \\
                ia-crime-moreno &   829 &                3.6 &  0.00 \\
\bottomrule
\end{tabular}